\documentstyle[11pt,newpasp,twoside,epsf]{article}
\def\edcomment#1{\iffalse\marginpar{\raggedright\sl#1\/}\else\relax\fi}
\reversemarginpar

\begin{document}
\title{Planetary Nebulae in the Magellanic Clouds: Probing Stellar Evolution and Populations}
 \author{Letizia Stanghellini}
\affil{Space Telescope Science Institute, 3700 San Martin Drive, Baltimore (MD) 21218; affiliated
with the Astrophysical Division, Space Science department of ESA.}

\begin{abstract}

This review contains: (1) the scientific 
motivations for studying Planetary Nebulae in the 
Magellanic Clouds; (2) a review of
this field of study, from the origins 
to the most recent results, focusing on 
the papers that have been published since the 
last IAU Symposium on Planetary Nebulae; 
(3) a review of the Hubble contribution to the field, from
the early results to 
our own Magellanic Cloud Planetary Nebula
program.

\end{abstract}
\section{Introduction}
Planetary Nebulae (PNs) are the relics of intermediate-mass star evolution, 
and 
they are important probes of stellar evolution, stellar
Populations, and cosmic recycling. PNs have been observed
in most galaxies (the Milky Way,
the Local Group galaxies, and distant galaxies up to 10 Mpc)
and in the medium between galaxies, thus they
also probe evolution and Populations in relation to their environment.

The success of the studies of Galactic PNs is based on their
proximity.
The details of the observations typically surpass the 
details of stellar and hydrodynamic models. Yet, the 
distance scale of Galactic PNs is uncertain to such a degree that 
the meaning of the comparison between observations and theory is hindered. 
By the same token, statistical studies 
of PN populations in the Galaxy suffer for the observational
bias against the detection of Galactic disk PNs, and for the patchy interstellar
extinction. Thus, while it is certainly wise to cherish the level of detail
offered by Galactic PN observations, we should also consider studying
PNs in other
galaxies, to approach the final phases of intermediate star evolution in a 
quantitative, statistical way.

PNs in the Magellanic Clouds (LMC, SMC), hundreds of low-extinction 
planetaries at uniformly known distances, are a real bounty for the stellar
evolution scientist. Moreover, the composition gradients between the LMC, the SMC,
and the Galaxy make studying MC PNs essential to understand the effects
of environment metallicity on their evolution. Finally,
understanding how stellar evolution operates in very low metallicity
ambients, such as the SMC, is a useful 
step toward the understanding of stellar Populations in distant galaxies.

In essence, I believe that the MCs are ideal laboratories to 
shed light on the still many unsolved
issues of Planetary Nebula formation and evolution.
Together with the detailed studies of individual PNs in the Galaxy, 
the comparison of large and homogeneous samples of the observed
properties of LMC and SMC PNs with the correct evolutionary models
will enormously improve our knowledge of the evolution of 
intermediate-mass stars, their stellar and nebular remnants, their
origin, and their interaction with the environment.

\section{Magellanic Cloud Planetary Nebulae: a field in evolution}

\subsection{Pioneers in the field}

PN studies in the MCs are relatively recent. 
The first discovery paper of MC PNs (Lindsay 
1955) contains the spectroscopic identification
of 17 SMC PNs. Very rapidly, studies of MC PN samples became 
common, and their importance evident
(e.g., Aller 1961; Westerlund 1964). 
Ground based observation of MC PN suffer from the fact that they are point 
sources, thus the contributions of nebular and stellar radiation are 
superimposed. Attempts to measure
the central stars magnitudes were hampered by the difficulties in separating
stellar and nebular contributions (e.g., Webster 1969).

From stellar evolution theory, we 
expect that some heavy elements (argon, neon, sulfur)
are not processed by stars in the mass range of the PN progenitors, 
thus the abundances of these elements
are the signature of the chemical mix of the
environment when the stellar progenitors were born. 
On the other hand, carbon, oxygen and 
nitrogen are at variation during the evolutionary paths that preceded the 
PN phases, and their relative abundances
probe the nuclear processes within the evolving star, the occurrence
of the-dredge up processes,
and the mass of the progenitors (Iben \& Renzini 1983). 
Since the central stars of MC PNs are not 
directly observable from the ground, the correct observations and analysis
of the nebular chemistry offers best insight of stellar evolution from
the nebular properties.

Observations with the IUE, combined with the optical spectral data acquired from the 
ground, allowed the abundance analysis of MC PNs. 
Space observations in the UV range was used for
the detection of the complete set of carbon lines at various ionization stages, 
and made the carbon abundance derivation much more reliable. 
The key results in abundance studies can be found, to name 
a few, in Peimbert (1984), Boroson \& Liebert (1989); Kaler \& Jacoby 
(1991). Optical spectroscopy of large samples of MC PNs
have been carried out by Dopita and collaborators (Meatheringham \& Dopita 
1991ab, Vassiliadis et al. 1992).
IUE observations were also used to measure the stellar luminosity beyond
the Lyman limit for the central stars, giving an estimate of the total stellar
luminosity, and and approximate estimate to the mass (Aller et al. 1987).
Several papers on MC PN spectroscopy, 
abundances, and the connection of nebulae and stellar evolution can be
found in the earlier IAU Symposia 
on Planetary Nebulae (e.g., Westerlund 1968; Feast 1968; 
Webster 1978), while the most recent, complete review on MC PNs
is due to Barlow (1989). 

Magellanic Cloud Planetary Nebula 
spectra are similar, in general, to the well-known Galactic ones.
In the MCs, as well as in the Galaxy, PNs have been
classified on the basis of their chemical content (see Peimbert 1978, 1997). 
New stellar models 
models with the MC  metallicity were build (Vassiliadis \& Wood 1993, 1994)
to improve the overall knowledge of the origin of the MC PNs.

\subsection{MC PN science since the last IAU Symposium}

Magellanic Cloud PNs become even more relevant in the recent years. Since the last
IAU Symposium on Planetary Nebulae, more than 70 papers related to MC PNs have 
been published, of the 363 papers ever published in this field \footnote{
These numbers are derived from an ADS search, and may not be complete}.
The most notable recent results are in two
major areas: The catalogs and emission line surveys, and
the nebular-stellar evolution connection. 

To date, 277 LMC PNs (Leisy et al. 1997) and 55 SMC PNs (Meyssonnier \& Azzoppardi
1993) are known. The total number of MC PNs have more than doubled
from the last count by Barlow (1989).
In the last few years, several emission line surveys have been completed,
or are near completion (e.g., UKST survey:
Morgan 1998, Parker \& Phillips 1998; UM/CTIO survey: 
Smith et al. 1996). 
Future analysis of these surveys is
essential for the future health of MC PN research. We expect that the PN counts 
in the MCs will increase significantly, improving the statistics of these studies.
For example,
Murphy \& Bessel (2000) found 107 new SMC PN candidates, their PN status remains
to be confirmed with spectroscopy.
One important aspect of these surveys is the
discovery of fainter PNs, that contributes to
increasing the reliability of the faint end of the MC PN 
luminosity function (see Jacoby, this volume), and to enlarge the pool of 
known evolved PNs.

Related to the populations of MC PNs is the 2MASS survey (Egan, Van Dyk, \& Price 2001). The 
importance of this multi-wavelength infrared survey to LMC PNs is
related to the spatial distribution of the different types of AGB stars. 
Egan et al. showed that
low mass AGB stars occupy the whole of the 
LMC projected volume, while the higher mass, younger Population, AGB stars populate 
preferentially the LMC bar. Chemical and morphological studies of large LMC PN samples 
should take these distributions into account, to relate the PN populations in the 
LMC to their immediate evolutionary progenitors.

Studies of the chemical content of MC PNs 
have been active in the last five years as well.
On the observational side, Leisy \& Dennefeld (1996), Costa, de Freitas Pacheco,
\& Idiart (2000), and Idiart \& Costa (this volume) have produced new chemical 
abundances 
for several MC PNs from optical and UV observations, enriching the databases
for studies on the dredge-up of post-AGB stars and on the ISM enrichment in galaxies.
On the theoretical side, van 
den Hoek \& Groenewegen (1997) calculated new chemical yields of the 
enrichment of the interstellar medium from synthetic evolution of intermediate-mass 
stars. With models from a wide range of initial masses and metallicities, 
including the 
MC metallicities, van 
den Hoek \& Groenewegen (1997) confirm that the yields of nitrogen and carbon 
change abruptly for masses higher than about 4 solar masses,  independent on
initial composition. 

\section{The Hubble Space Telescope contribution to the field}

\subsection{Early Hubble observations of Magellanic Cloud Planetary Nebulae}

Extended studies of Galactic PNs have shown that PN morphology 
is intimately related to the mass and evolution of their central stars,
to their stellar progenitors, and to the nebular chemistry (e.g., Manchado,
this volume). Morphology appears then to be an essential PN property, to
be studied statistically, and to be compared case by case to
the nebular and the stellar properties. In the case of the LMC and
the SMC, morphological studies became possible with the use of the
cameras on board the Hubble Space Telescope. 
Hubble observations have also the capability of spatially
separate the image of the nebula and that of the central star,
making direct stellar analysis possible.

The early narrow-band images of 
MC PNs were obtained before the first Hubble servicing mission 
(i.e., before the 
installation of COSTAR on Hubble) with the {\it Faint Object Camera} 
(Blades et al. 1992). 
Other images by Blades et al. have later been published by 
Stanghellini et al. (1999),
where the quality of the pre-COSTAR images was validate through 
their comparison with post-COSTAR images of the same objects. 
These papers have made available
15 MC PN images usable for statistical and morphological studies, while another 15 LMC 
PNs have been observed with the {\it Planetary Camera 1} by Dopita et al. (Dopita et al. 1996; 
Vassiliadis et al. 1998a). Finally, an additional ten {\it Wide Field and Planetary Camera 2}
narrow-band images of LMC PNs are available in the 
Hubble Data Archive (program 6407, PI: Mike Dopita).

The UV and optical spectroscopic capability of Hubble have been
employed to greatly improving the quality of the spectrophotometric 
calibration of SMC and LMC PNs.
Among the many
papers that have been published based on Hubble spectroscopy
(e.g., Vassiliadis et al. 1998b, Dopita et al. 1997ab),
two results from Dopita et al. are worth noting:
first, it is shown
that there are different populations of PNs within 
the LMC, and that is it 
possible to discern these populations on the basis of their alpha-element abundance (see 
Fig. 3 in Dopita et al. 1997b). Second, by relating the expansion of 
the LMC PNs 
to the central star position on the HR diagram, it is inferred that 
several central stars of LMC PNs sustain their evolution via helium-burning, rather than 
hydrogen-burning. Even if the latter result is hindered by 
the obvious uncertainties on the location of the stars in the HR 
diagram (the stars were not directly observed in these narrow-band 
images, and their location on the HR diagram
was inferred from nebular analysis), they are still important as they hint
to a possible
dichotomy of PN populations in the LMC.

\subsection{Our program, based on Hubble data, and the future}

Statistical studies of MC PNs and their central stars
need an homogeneous, large imaging databases to achieve significance.
In the past two years, a project with HST was implemented
by myself and my collaborators,
with the aim of obtaining morphology, 
nebular physics, and stellar parameters, for all the
known PNs in the MCs. The scientific rationale is to
derive a complete description of formation and evolution of 
PNs according to their environment, based on the 
confrontation of the HST and ancillary ground-based
data with the stellar and hydrodynamic models of PN evolution. 

The observing strategy invokes slitless spectroscopy 
with the {\it Space Telescope Imaging
Spectrograph} (STIS), coupled with
STIS broad-band imagery. This technique provides pseudo narrow-band
images in the major emission lines (e.g., H-alpha, H-beta, 
[O III] 5007 A, [N II] 6584 A, [S II] 6716-6731 A),
suitable to derive the size, 
the ionization stage, the gas density, the extinction,
and the morphology of the nebulae, plus broad-band images to determine
the magnitude of the central stars. The observations, 
performed in snapshot mode, gather morphological
and stellar information in less than one orbit for each MC PN, 
making it a cost-effective HST program.

The data analysis plan is at an early stage, yet we have already obtained 
important results from the samples of MC PNs, to add to the
value of
the sheer quantity and quality of the images collected so far.
The major results from the observations and analysis
can be summarized as follows:

\begin{figure}
\plotone{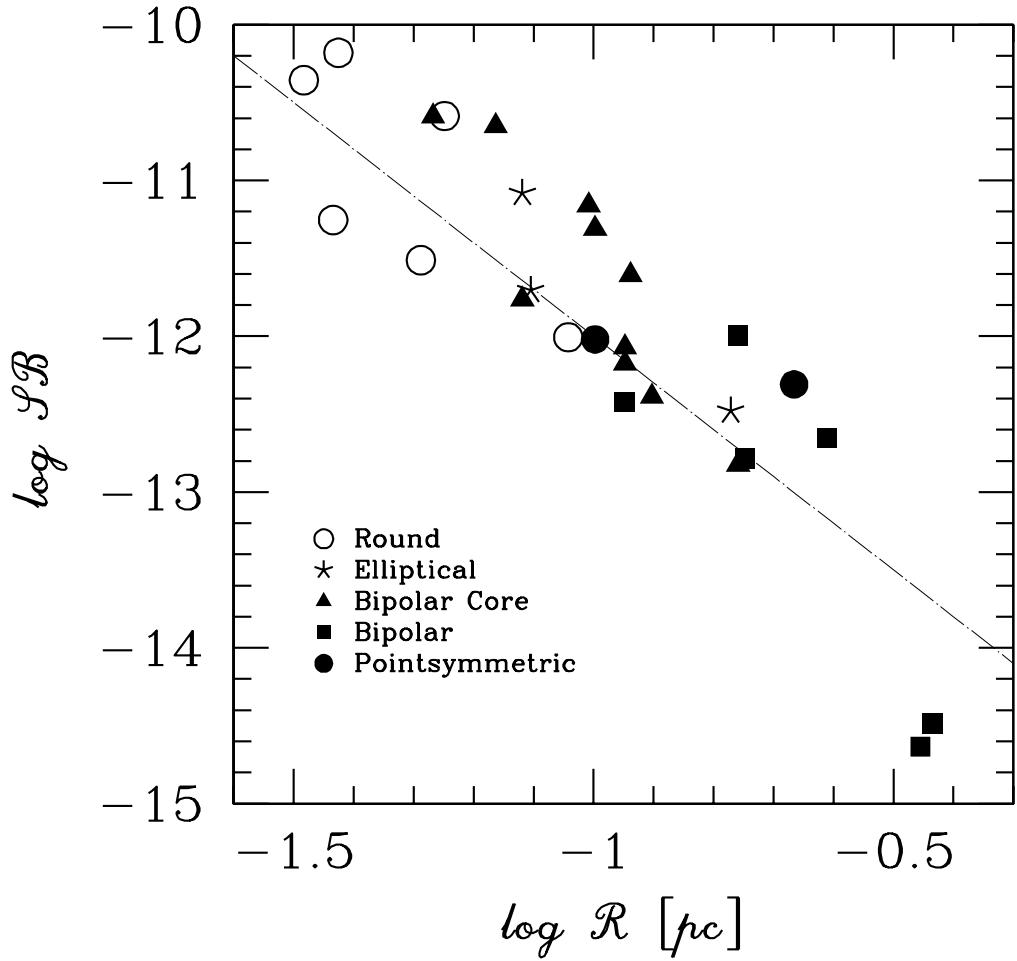}
\caption{The [O III] 5007 logarithmic surface brightness versus the 
logarithmic (physical) photometric radii, 
for LMC Planetary Nebulae.
The thin line represents a rough fit to the 
data, SB ~ R**-3.}
\end{figure}

\begin{enumerate}
\item{
PNs in the Clouds have substantially the same morphologies
than Galactic PNs: Symmetric (Round and Elliptical) and
Asymmetric (Bipolar, Bipolar Core, and Pointsymmetric) PNs were observed.
There is a larger fraction of Asymmetric to Symmetric PNs 
in the LMC than in the Galaxy; since most Asymmetric Galactic 
PNs are Galactic disk objects, the result may simply disclose
the selection effects that play against the detection of Galactic disk PNs. 
The fraction of Symmetric to Asymmetric PNs in the LMC is different that 
that of the SMC, indicating different mixes of PN populations across 
galaxies.}

\item{
The surface brightness of the LMC 
PNs in the light of [O III] 5007 A, correlates with the photometric
radii of the nebulae, as shown in Figure 1. 
The photometric radius of a PN roughly traces the rate of its dynamic
evolution (velocities of PNs are within a very narrow range, see
Shaw et al. 2001), 
therefore the relation of Figure 1 can be used to 
constraint the hydrodynamic models. In fact, preliminary models 
show a similar surface brightness decline in PN evolution (Villaver
et al. in preparation).
In Figure 1 the symbols indicate the PN morphology, as listed in
the legend.
We infer that Asymmetric 
PNs have a more rapid 
dynamic evolution than Round PNs. The
distribution of the SMC PNs on the plane of Figure according to
 morphological type
is similar to that, shown, of LMC PNs. We do not plot the SMC PNs in Figure 1,
since their calibrated surface brightness
are yet to be measured from the Hubble spectra.}

\item{
We found that asymmetric PNs are neon and sulfur rich, compared
to Round and Elliptical PNs, in the LMC (Stanghellini et al., 2000). 
This result is broadly consistent with the predictions 
of stellar evolution if the progenitors 
of Asymmetric PNs have on average larger masses than the progenitors of 
Symmetric PNs, independent on assumptions, or relation to, a possible
stellar multiplicity of the
progenitors. This result is also in broad agreement with
the findings in Galactic PNs that bipolar Planetary Nebulae
have more massive central stars (e.g., Stanghellini, Corradi, \&
Schwarz 1993), and 
bears on the question of formation mechanisms 
for asymmetric PNs, specifically, that the genesis of PNs structure should 
relate strongly to the Population type, and by inference the mass, of the 
progenitor star, and less strongly on whether the central star is a member 
of a close binary system.}

 \end{enumerate}

An important product of our program is the prepared data set, 
available within
the HST Data Archive. The observations, images, analysis,
publications, and other useful links for PN science can be found at:

{\bf http://archive.stsci.edu/hst/mcpn/}

In the future, we plan to furtherly exploit the data that we are collecting.
One of our aims is
the determination of the central star photometry, and the
direct calculation of the stellar luminosity. By deriving the effective 
temperature via Zanstra analysis, or with the analysis of the stellar 
spectra, we will 
be able to locate the central stars in the logL-logT$_{\rm eff}$ plane. 
The comparison of these locations with the evolutionary tracks will lead to the
determination of the central star's masses. The stellar properties 
will then be related to the nebular ones, to disclose all the possible relation 
between stars and nebulae, with the aim of finding direction toward the
understanding of PN formation and evolution.
The stellar properties will also be studied with the UV STIS slitless
spectra that we are presently acquiring.
The stellar data will constraint the
the detailed hydrodynamic models, that we will build specifically 
for MC PNs to compare them to the Hubble images. 
Optical and ultraviolet spectra will be used to determine the
PN abundances, with the aim of confirming our results on LMC PN 
progenitors.

\section{Acknowledgements}

Sun Kwok and Micheal Dopita are warmly thanked for inviting me to review this 
important subject. Thanks to the organizers, and in particular to Peter Wood, for their 
help in solving a number of questions. Section 3.2 is based on work
in collaboration with Richard Shaw, Max Mutchler, Bruce Balick, Chris Blades, and
Stacy Palen.

\end{document}